\documentclass[fleqn,10pt]{wlscirep}
\usepackage{bbold}
\newcommand{\bra}[1]{\ensuremath{\left\langle#1\right|}}
\newcommand{\ket}[1]{\ensuremath{\left|#1\right\rangle}}

\title{Supervised Quantum Learning without Measurements}

\author[1,*]{Unai Alvarez-Rodriguez}
\author[1]{Lucas Lamata}
\author[2]{Pablo Escandell-Montero}
\author[2]{Jos\'{e}~D.~Mart\'{i}n-Guerrero}
\author[1,3]{Enrique Solano}
\affil[1]{Department of Physical Chemistry, University of the Basque Country UPV/EHU, Apartado 644, 48080 Bilbao, Spain}
\affil[2]{IDAL, Electronic Engineering Department, University of Valencia, Avgda. Universitat s/n, 46100 Burjassot, Valencia, Spain}
\affil[3]{IKERBASQUE, Basque Foundation for Science, Maria Diaz de Haro 3, 48013 Bilbao, Spain}
\affil[*]{Correspondence to unaialvarezr@gmail.com}

\begin{abstract}
We propose a quantum machine learning algorithm for efficiently solving a class of problems encoded in quantum controlled unitary operations. The central physical mechanism of the protocol is the iteration of a quantum time-delayed equation that introduces feedback in the dynamics and eliminates the necessity of intermediate measurements. The performance of the quantum algorithm is analyzed by comparing the results obtained in numerical simulations with the outcome of classical machine learning methods for the same problem. The use of time-delayed equations enhances the toolbox of the field of quantum machine learning, which may enable unprecedented applications in quantum technologies.
\end{abstract}

\begin{document}

\flushbottom
\maketitle
\thispagestyle{empty}

\section*{Introduction}
One of the main consequences of the revolution in computation sciences, started by Alan Turing, Konrad Zuse and John Von Neumann, among others~\cite{Turing,ComputationBook}, is that computers are capable of substituting us and improving our performance in an increasing number of tasks. This is due to the advances in the development of complex algorithms and the technological refinement allowing for faster processing and larger storage. One of the goals in this area, in the frame of bio-inspired technologies, is the design of algorithms that provide computers human-like capacities such as image and speech recognition, as well as preliminary steps in some aspects related to creativity. These achievements would enable us to interact with computers in a more efficient manner. This research, together with other similar projects, is carried out in the field of artificial intelligence~\cite{arin}. In particular, researchers in the area of machine learning (ML) inside artificial intelligence are devoted to the design of algorithms responsible of training the machine with data, such that it is able to find a given optimal relation according to specified criteria \cite{Alpaydin}. More precisely, ML is divided in three main lines depending on the nature of the protocol. In supervised learning, the goal is to teach the machine a known function without explicitly introducing it in its code. In unsupervised learning, the goal is that the machine develops the ability to classify data by grouping it in different subsets depending on its characteristics. In reinforcement learning, the goal is that the machine selects a sequence of actions depending on its interaction with an environment for an optimal transition from the initial to the final state. 

The previous ML techniques have also been studied in the quantum regime in a field called quantum machine learning~\cite{hb,qml1,qml2,qmlLloydReview,wittek,qml3,qml4,qml5} with two main motivations. The first one is to exploit the promised speedup of quantum protocols for improving the already existing classical ones. The second one is to develop unique quantum machine learning protocols for combining them with other quantum computational tasks. Apart from quantum machine learning, fields like quantum neural networks, or the more general quantum artificial intelligence, have also addressed similar problems \cite{qnn,qnn1,qnn01,qnn14,qnn16}. 

Here, we introduce a quantum machine learning algorithm for finding the optimal control state of a multitask controlled unitary operation. It is based on a sequentially-applied time-delayed equation that allows one to implement feedback-driven dynamics without the need of intermediate measurements. The purely quantum encoding permits to speedup the training process by evaluating all possible choices in parallel. Finally, we analyze the performance of the algorithm comparing the ideal solution with the one obtained by the algorithm. 

\section*{Results}
\subsection*{Quantum Machine Learning Algorithm}
The first step in the description of the algorithm is the definition of the concept of multitask controlled unitary operations $U$. In essence, these do not differ from ordinary controlled operations, but the multitask label is selected to emphasize that more than two operations are in principle possible. $U$ acts on $\ket{\psi}$, being $\ket{\psi} \in \mathbb{C}^d \otimes \mathbb{C}^d$, a quantum state belonging to the tensor product of the control, $\mathcal{H}_c \subset \mathbb{C}^d$, and target, $\mathcal{H}_t \subset \mathbb{C}^d$,  Hilbert spaces. The dimension of both subspaces is the same, $d$, and depends on the particular problem to address. Mathematically, we define $U$ as
\begin{equation}
U=\sum^{d}_{i=1}\ket{c_i}\bra{c_i}\otimes s_i,
\end{equation}
where $\ket{c_i}$ denotes the control state, and $s_i$ is the reduced or effective unitary operation that $U$ performs on the target subspace when the control is on $\ket{c_i}$. 

The goal of our algorithm is to explore the control subspace $\mathcal{H}_c$ and find the control state that maximizes the implementation of a known $s$, $s : \mathcal{H}_t \rightarrow \mathcal{H}_t$, which is given in terms of the $\ket{\text{in}}$ and $\ket{\text{out}}$ states as $s \ket{\text{in}} = \ket{\text{out}}$. Therefore, our algorithm is appropriate when $U$ is experimentally implementable but its internal structure, the relation between $\ket{c_i}$ and $s_i$, is unknown. In other words, our algorithm enables the training of the control subspace $\mathcal{H}_c$ by providing data about the target subspace $\mathcal{H}_t$, in order to achieve that the complete system implements the desired $s$ operation in the target subspace $\mathcal{H}_t$. Our inspirations for the model of controlled unitary operations are supervised learning protocols, in which the goal is that the system is able to learn a given known function. Here, the control subspace plays the role of the memory of the system. This control, or memory, is the mechanism by which the system is able to store the information about the operation that it has to implement.  The idea of our algorithm is that the user transmits the information of the operation the system has to make. Therefore, the goal is not to perform a given gate, but to save this information in the system.

The protocol consists in sequentially reapplying the same dynamics in such a way that the initial state in the target subspace is always $\ket{in}$, while the initial state in the control subspace is the output of the previous cycle. The equation modeling the dynamics is
\begin{equation}
\frac{d}{dt} \ket{\psi(t)} = -i \left[  \theta(t-t_i)\theta(t_f -t)\kappa_1 H_1 \ket{\psi(t)} + \kappa_2 H_2 \left(  \ket{\psi(t)} - \ket{\psi(t-\delta)} \right)  \right].
\label{epv}
\end{equation}
In this equation $\theta$ is the Heaviside function, $H_1$ is the Hamiltonian giving rise to $U$ with $U=e^{-i \kappa_{1} H_{1}(t_f -t_i) }$, and $H_2$ is the Hamiltonian connecting the input and output states, with $\kappa_1$ and $\kappa_2$ the coupling constants of each Hamiltonian. We point out that this evolution cannot be realized with ordinary unitary or dissipative techniques. Nevertheless, recent studies in time delayed equations provide all the ingredients for the implementation of this kind of processes~\cite{grimsmo,whalen,fede,are}. Up to future experimental analyses involving the scalability of the presented examples, the inclusion of time delayed terms in the evolution equation is a realistic approach in the technological framework provided by current quantum platforms. Another important feature of Eq. \eqref{epv}, which is related with the delayed term, is that it only acquires physical meaning once the output is normalized.

Regarding the behavior of the equation, each term has a specific role in the learning algorithm. The mechanism is inspired in the most intuitive classical technique for solving this problem, which is the comparison between the input and output states together with the correspondent modification of the control state. Here, the first Hamiltonian produces $U$ while the second Hamiltonian produces the reward by populating the control states responsible of the desired modification of the target subspace. The structure of $H_2$ guarantees that only the population in the control $\ket{c_i}$ associated with the optimal $s_i$ is increased, 
\begin{equation}
H_2=\mathbb{1}\otimes\left( -i \ket{\text{in}}\bra{\text{out}}+i\ket{\text{out}}\bra{\text{in}} \right).
\end{equation}
Notice that while this Hamiltonian does not contain explicit information about $\ket{c_i}$, the solution of the problem, its multiplication with the feedback term, $ \ket{\psi(t)} - \ket{\psi(t-\delta)}$, is responsible for introducing the reward as an intrinsic part of the dynamics. This is a convenient approach because it eliminates the measurements required during the training phase. In this case where we employ a single pair of $\{ \ket{\text{in}}$, $\ket{\text{out}}$\} target states, $H_2$ is fixed and time independent. However, this could change in a more complex situation of $p$ pairs of $\{\ket{\text{in}}$, $\ket{\text{out}}\}$ target states, such that $s=\sum^{p}_{j} \ket{\text{out}}_j \otimes \bra{\text{in}}_j $, where $H_2$ would also be time independent but different in each episode. Even if this generalization is not included in this article, it points out a promising direction for enhancing the protocol.

We would also like to remark the similarity existing between the effect of the delay term in our quantum evolution and gradient ascent techniques in algorithms for optimization problems~\cite{arin}. A possible strategy to perform the learning protocol would be to feed the system with random control states, measure each result, and combine them to obtain the final solution. However, we have discovered that it suffices to initialize the control subspace in a superposition of the elements of the basis. We would like to remark that this purely quantum feature reduces significantly the required resources, because a single initial state replaces a set of random states large enough to cover all possible solutions. 

\subsection*{Numerical Simulations}
We have numerically tested our proposed algorithm in a selection of examples covering the cases with unique or multiple solutions, as well as higher-dimensional systems. We consider as a figure of merit the fidelity function given by the trace of the product between the control state obtained by the algorithm and the ideal control state. In order to recover the solution of the problem we need to trace out the target degrees of freedom, obtaining a density matrix. Therefore, the iteration of the protocol would require solving Eq. \eqref{epv} written for density matrices. This turns out to be a nontrivial task given the non-local cross terms of the generalized master equation, that reads,
\begin{align}
\frac{d}{dt} \ket{\psi(t)}\bra{\psi(t)}=&-i \left[ \theta(t-t_i)\theta(t_f -t)\kappa_1 H_1 + \kappa_2 H_2, \ket{\psi(t)}\bra{\psi(t)}\right] \nonumber \\& +i \kappa_2(H_2 \ket{\psi(t-\delta)}\bra{\psi(t)}-\ket{\psi(t)}\bra{\psi(t-\delta)}H_2).
\label{epm}
\end{align}
To achieve the solution in the most efficient way, we have decomposed each density matrix in a convex sum of pure states and solved the vector equation in Eq.~(\ref{epv}) for each of them separately, retrieving the total solution as a linear convex superposition of the individual ones. This method is consistent due to the linearity of Eq. \eqref{epm}.

\begin{figure}[t]
\includegraphics[width=0.66\textwidth]{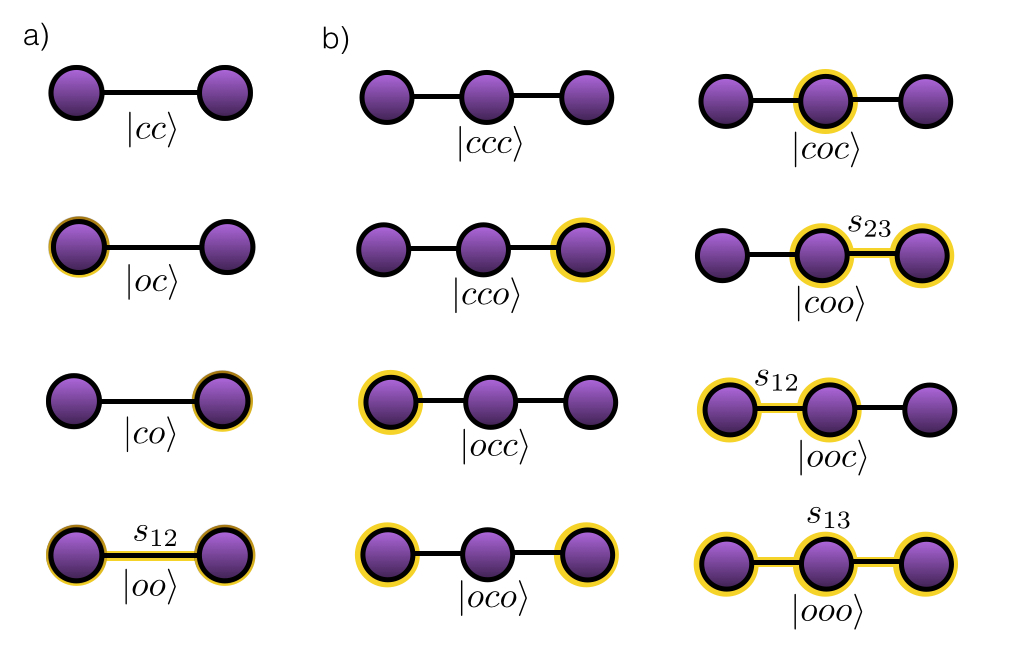}
\caption{{\bf Node line networks.} We plot the graphical representation of every control state in the two, a), and three, b), node line networks. The circles around the nodes denote the control being in the open state. The effective operation that the control performs on the target subspace is the $s_{ij}$ SWAP gate between nodes $i$ and $j$.}
\label{fl1}
\end{figure}

\subsubsection*{Definition of the SWAP gate problem}
A first specific example we address in this manuscript is given by the excitation transport produced by the controlled SWAP gate. In this scenario, the complete system is an $n$-node network, where each node is composed by a control and a target qubit. Therefore, the control and target subspaces are defined as $\mathcal{H}_c \subset (\mathbb{C}^2)^{\otimes n}$ and $\mathcal{H}_t \subset (\mathbb{C}^2)^{\otimes n}$. The excitations in this system belong to the target subspace and are exchanged between two nodes, when both nodes are in a particular state of the control subspace. The control states are in a superposition of open and close, $\ket{o}$ and $\ket{c}$, while the target qubits are written in the standard $\{ \ket{0}, \ket{1} \}$ basis denoting the absence or presence of excitations. We define $U$, the multitask controlled unitary operation, to implement the SWAP gate between connected nodes only if all the controls of the corresponding nodes are in the open state, $\ket{o}$. See Fig.~\ref{fl1} for a graphical representation of the most simple cases, the two and three node line networks. The explicit formula for $U_2$ is given by
\begin{align}
U_2= & ( \ket{cc}\bra{cc}+\ket{co}\bra{co}+\ket{oc}\bra{oc} ) \otimes \mathbb{1} + \ket{oo}\bra{oo}\otimes s_{12}
\end{align}
where $s_{ij}$ represents the SWAP gate between qubits $i$ and $j$. Here, the first two qubits represent the control subspace and the last two represent the target subspace. Although we have employed unitary operations for illustration purposes, the equation requires the translation to Hamiltonians. In order to do so, we first select $\kappa_1(t_f - t_i)$ to be $\pi/2$ and calculate the matrix logarithm, which yields the result for $H_1$ in Eq. \eqref{epv}, $H_1=(\ket{oo}\bra{oo})\otimes h_{12}$. Denoting with $\sigma_k$ the Pauli matrices, $h_{ij}$ for $i<j$ reads
\begin{equation}
h_{ij}=\frac{1}{2}\left(\sum^{3}_{k=1}\mathbb{1}^{\otimes i-1}\otimes \sigma_k \otimes \mathbb{1}^{\otimes j-i-1}\otimes \sigma_k\otimes\mathbb{1}^{\otimes n-j} -\mathbb{1}^{\otimes n} \right).
\end{equation}

\begin{figure}[h]
\includegraphics[width=1.03\textwidth]{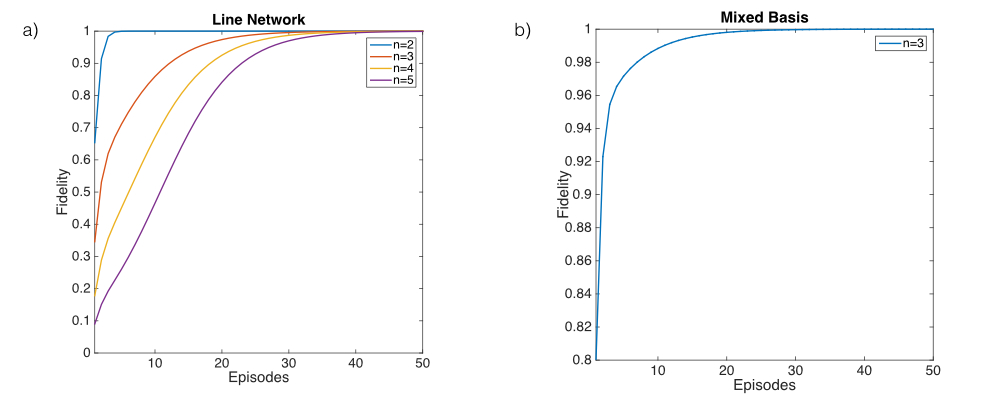}
\caption{{\bf Learning curves for single solutions.} a) We plot the fidelity of the learning process as a function of the number of episodes for the first examples of $n$-node line networks. We have selected the open state, $\ket{o}=\ket{1}$ of the $\{\ket{0}, \ket{1}\}$ basis. b) We plot the fidelity for a different selection of $\ket{o}$ in the $n=3$ case. Here, we have rotated the control states with the goal of testing the algorithm for an arbitrary basis. The solution, $\ket{ooo}$, is given by $\frac{1}{\sqrt{2}}[\ket{0}+\ket{1}]\otimes\ket{1}\otimes[\cos{(\pi/3)}\ket{0}+\sin{(\pi/3)}\ket{1}]$. }
\label{fl2}
\end{figure}

\subsubsection*{Unique solution of the quantum machine learning algorithm} The first family of problems we address is {\it n-node line networks}, in which the nodes are located in a unidimensional array and are only connected with their closest neighbors. The goal is to find the control state that allows transmitting an excitation from the first to the last node of the network, which in this case requires that all intermediate connections are active. The pair of $\{ \ket{in}, \ket{out} \}$ is determined by these constrains as $\ket{\text{in}}=\ket{1}\ket{0}^{\otimes n-1}$ and $\ket{\text{out}}=\ket{0}^{\otimes n-1}\ket{1}$. Accordingly the problem has a unique solution, given by the control state with all the nodes open, $\ket{o}^{\otimes n}$. The parameters we have selected are $\delta=1$, $\kappa_1=100$, $\kappa_2=10$ and $T=2$, where $T$ represents the total duration of each episode. In Fig.~\ref{fl2} we plot the results together with the required resources. These examples show how the algorithm is properly working for this family of problems independently of the natural basis of $U$. The $H_1$ Hamiltonians employed in the simulations for $n=2,3,4$ are given by
\begin{align}
H^{2}_1=&|oo\rangle\langle oo|\otimes h_{12}, \nonumber \\ 
H^{3}_1=&|ooo\rangle\langle ooo|\otimes h_{13} + |ooc\rangle\langle ooc|\otimes h_{12} + |coo\rangle\langle coo| \otimes h_{23}, \nonumber \\
H^{4}_1=&|oooo\rangle\langle oooo| \otimes h_{14} + |oooc\rangle\langle oooc|\otimes h_{13} + ( |oocc\rangle\langle oocc| + |ooco\rangle\langle ooco| ) \otimes h_{12} \nonumber  \\ &+ (|ccoo\rangle\langle ccoo|+|ocoo\rangle\langle ocoo|)\otimes h_{34} + |cooc\rangle\langle cooc|\otimes h_{23} + |cooo\rangle\langle cooo| \otimes h_{24}. 
\end{align}

\begin{figure}[h]
\includegraphics[width=0.66\textwidth]{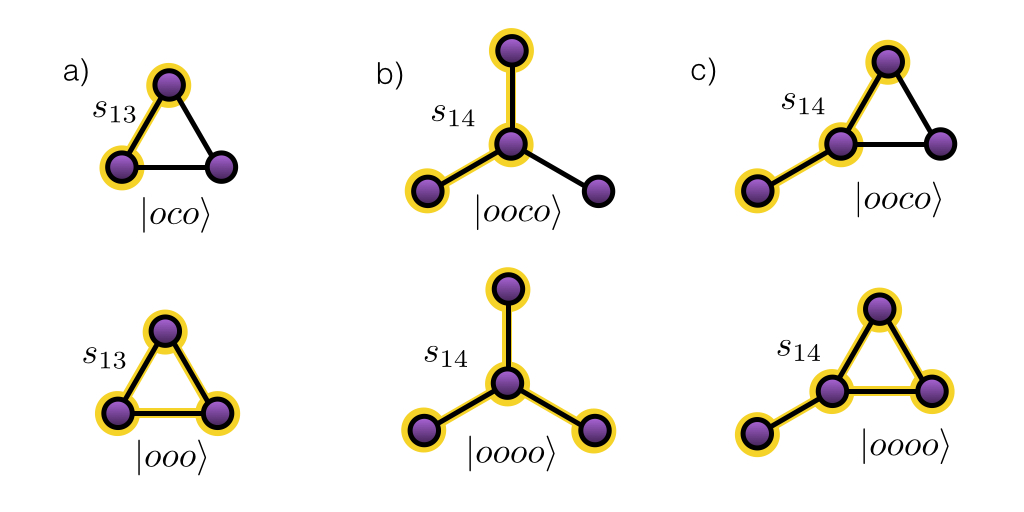}
\caption{{\bf Networks with two solutions.} We schematize the A, B, and C networks in a), b) and c), respectively. In each of them, we write the pair of solution control states that corresponds to the control performing the $s_{13}$ (a)  and $s_{14}$ (b,c) gates in the target subspace.}
\label{fl3}
\end{figure}

\subsubsection*{Multiple solutions of the quantum machine learning algorithm} 
\begin{figure}[h]
\includegraphics[width=\textwidth]{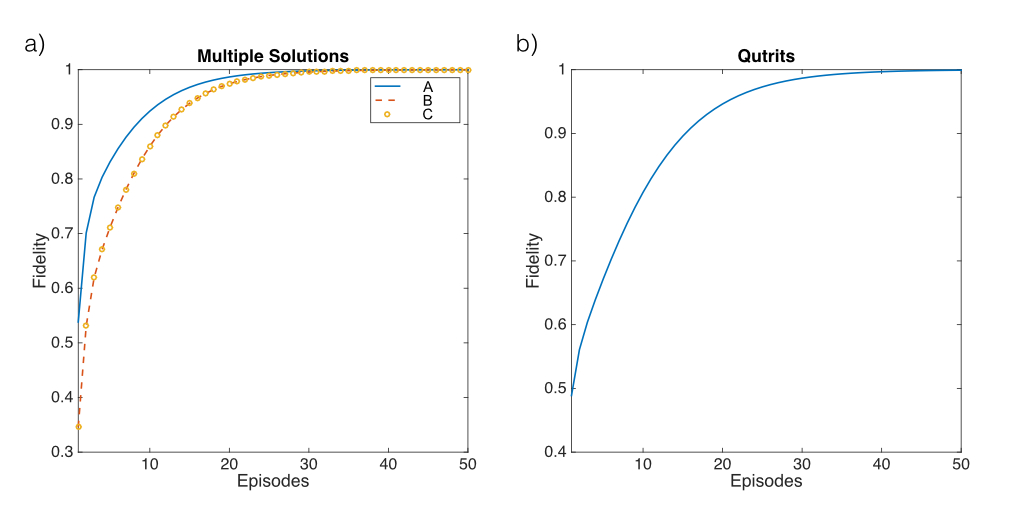}
\caption{{\bf Learning curves for two solutions and qutrit problems.} a) We depict the learning curve for the A, B, and C networks as a function of the number of episodes. Notice that the curves for the B and C networks are identical. b) We depict the learning curve for the multitask controlled unitary operation acting on two qutrits as a function of the number of episodes. Here, $\ket{\text{in}}=\ket{0}$, $\ket{\text{out}}=\ket{2}$ and the solution is given by $\ket{c_2}=\ket{1}$, where the control states coincide with the basis of the qutrit space.}
\label{fl4}
\end{figure}

We address now a set of more complicated networks which will allow us to clarify how the algorithm performs when solving problems with multiple solutions. These are the A network for three nodes and the B and C networks for four nodes, depicted in Fig.~\ref{fl3}. The goal of the algorithm is the same as in the previous case, i.e., to find the control state able of sending an excitation from the first to the last node. The difference is that these networks accept two pure states and their superpositions as solutions, a feature that is reflected in the result obtained with the algorithm. The asymptotic state achieved under the feedback induced quantum learning equation is a quantum superposition of both solutions, see Fig.~\ref{fl4}a for the numerical simulations. In this case, the previous definition of the fidelity is not valid. Therefore, we provide a new one in terms of the \ket{\text{in}} and \ket{\text{out}} states of the target space and the Hamiltonian $H_1$. The new fidelity corresponds to the trace of the product between the ideal output \ket{\text{out}}, and the output obtained with the control state achieved by the algorithm after acting on \ket{\text{in}}. Both ideal and real outputs belong to the target subspace. While the \{\ket{\text{in}}, \ket{\text{out}}\} pair is the same as in the previous case, the $H_1$ Hamiltonians change their definition to
\begin{align}
H^{A}_{1}=&(|ooo\rangle\langle ooo|+|oco\rangle\langle oco|) \otimes h_{13} + |ooc\rangle\langle ooc| \otimes h_{12} + |coo\rangle\langle coo| \otimes h_{23}, \nonumber \\
H^{B}_{1}=&(|oooo\rangle\langle oooo|+|ooco\rangle\langle ooco|)\otimes h_{14} + |cooo\rangle\langle cooo|\otimes h_{34} + |oooc\rangle\langle oooc| \otimes h_{13}  \nonumber \\ & + |oocc\rangle\langle oocc| \otimes h_{12} + |cooc\rangle\langle cooc| \otimes h_{23} + |coco\rangle\langle coco| \otimes h_{24}, \nonumber \\
H^{C}_{1}=&|oocc\rangle\langle oocc| \otimes h_{12} + |cooc\rangle\langle cooc| \otimes h_{23} + (|coco\rangle\langle coco| + |cooo\rangle\langle cooo|) \otimes h_{24} + |oooc\rangle\langle oooc| \otimes h_{13} \nonumber \\ & + (|ocoo\rangle\langle ocoo| + |ccoo\rangle\langle ccoo|  )\otimes h_{34} +  (|ooco\rangle\langle ooco| + |oooo\rangle\langle oooo|  )\otimes h_{14}.
\end{align}

For the cases studied, the complete set of solutions is obtained encoded in the outcome of the algorithm. This is convenient because it allows one to design a protocol to select a specific optimal solution according to given criteria. In the networks we are analyzing, one might want to obtain the most efficient solution, defining efficiency as achieving the transmission of the excitation while minimizing the number of open nodes. In order to accomplish this task a dissipative term has to be included in the evolution equation, in order to filter out the undesired solutions. We point out that a control-dependent dissipation affects the target subspace, modifying the protocol in the required manner. We explicitly write the Lindblad operators $\sigma_i$ and dissipation constants $\gamma_i$ for a two-node case, as follows
\begin{eqnarray}
\nonumber &&\sigma_1=\ket{co01}\bra{co11}+\ket{co00}\bra{co10}, \hspace{0.25cm} \sigma_2=\ket{co00}\bra{co01}+\ket{co10}\bra{co11}, \\ \nonumber &&\sigma_3=\ket{oc00}\bra{oc10}+\ket{oc01}\bra{oc11}, \hspace{0.25cm} \sigma_4=\ket{oc10}\bra{oc11}+\ket{oc00}\bra{oc01}, \\ \nonumber &&\sigma_5=\ket{oo00}\bra{oo10}+\ket{oo01}\bra{oo11}, \hspace{0.25cm} \sigma_6=\ket{oo00}\bra{oo01}+\ket{oo10}\bra{oo11}, \\ && \gamma_1=\gamma_2=\gamma_3=\gamma_4, \hspace{0.25cm} \gamma_5=\gamma_6=2\gamma_1.
\end{eqnarray} 
Instead of solving the master equation, we have employed the quantum jump formalism, which allows one to work with Eq.~\eqref{epv} instead of Eq.~\eqref{epm}, with the consequent simplicity. The dissipation can be modeled with an additional term $H_D = \frac{-i}{2} \sum \gamma_i \sigma^{\dag}_{i}\sigma_{i}$ in the first part of the time delayed equation in the absence of a decay event. Therefore, in order to assure that the non-Hermitian Hamiltonian accounts for the real evolution of the system, one has to properly balance the relation between $\kappa_1$ and $\gamma_i$. 
\begin{equation}
\frac{d}{dt} \ket{\psi(t)} = -i \left[  \theta(t-t_i)\theta(t_f -t)(\kappa_1 H_1 + H_D)\ket{\psi(t)} + \kappa_2 H_2 \left(  \ket{\psi(t)} - \ket{\psi(t-\delta)} \right)  \right].
\end{equation}
A non-dissipative alternative consists in the modification of the coupling constant associated with each of the control-target pairs in the unitary operation. These two techniques allow us to find the shortest path between two nodes in a network once the natural basis of the unitary is known. 

\subsubsection*{Extension to qudits} Another possible aspect to study is the extension of the algorithm to higher-dimensional building blocks. We provide an example in which the optimal control state for a multitask controlled unitary operation acting on qutrits is obtained. This operation $U$ is defined in terms of the control states $\ket{c_i}$ as
\begin{equation}  
U=\ket{c_1}\bra{c_1}\otimes \mathbb{1} + \ket{c_2}\bra{c_2}\otimes \left( \begin{array}{ccc} 0&1&0\\0&0&1\\1&0&0 \end{array} \right) +\ket{c_3}\bra{c_3}\otimes \left( \begin{array}{ccc} 0&0&1\\1&0&0\\0&1&0 \end{array} \right).
\end{equation}
where the first qutrit belongs to the control subspace and the second one belongs to the target subspace. Although no network is defined in this case, the goal of the algorithm is to find the control state that realizes the \ket{\text{in}}-\ket{\text{out}} transition in the target subspace. In this problem, the system consists of a single control qutrit and a single target qutrit. See Fig.~\ref{fl4} for a numerical simulation of the learning process in this particular case. 

\begin{figure}[h]
\includegraphics[width=0.5\textwidth]{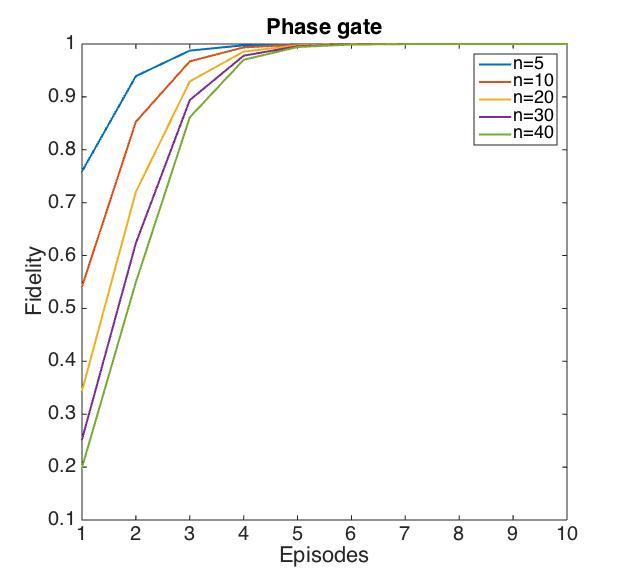}
\caption{{\bf Learning curves for phase gates.} a) We plot the fidelity of the learning process as a function of the number of episodes for the problem of finding the appropriate phase gate. }
\label{fl5}
\end{figure}

\subsubsection*{Extension to phase gates} All examples discussed until this point consisted on $s$ gates whose effect can be understood as a permutation of the basis elements. Let us consider now a different scenario in which the operations in the target subspace are phase gates, $s_i = \mathbb{1} -2 |i\rangle\langle i |$, therefore, the complete unitary operation reads $U=\sum |i\rangle\langle i|\otimes (\mathbb{1} - 2 |i\rangle\langle i |)$ with $i \in [1,n]$. If we choose the reference target states to be,
\begin{equation}
\ket{\text{in}}=\frac{1}{\sqrt{2}} (\ket{1}+\ket{n}),  \hspace{0.25cm} \ket{\text{out}}=\frac{1}{\sqrt{2}} (\ket{1}-\ket{n}),
\end{equation}
we know {\it a priori} that the only solution is given by $s=\mathbb{1}-2 |n\rangle\langle n|$ associated to control $\ket{c}=|n\rangle\langle n|$. We perform a numerical experiment to analyze how the initial equally weighted control state, $\frac{1}{\sqrt{n}} \sum \ket{i}$, converges to the solution under the action of $H_1=-2 \sum{\ket{i}\bra{i}\otimes\ket{i}\bra{i}}$ depending on the dimension of the system. See Fig.~\ref{fl5} for the simulations. The results show that our algorithm is particularly efficient for this selection of Hamiltonians, given that the solution is reached in $O(\sqrt{n})$ for all the cases studied.

\subsection*{Efficiency of the Quantum Machine Learning Algorithm}
It is important to mention that the simulations and techniques we provide here constitute an analysis of our quantum machine learning algorithm, but our aim in this work is not to demonstrate scalability or quantum speedup. It would be convenient to analytically solve Eq. \eqref{epv} in order to rigorously analyze the scope of the algorithm and be able to obtain information about its scalability for general problems. Since we have not solved the dynamics analytically, we evaluate the performance by comparing our results with the ones obtained via different methods. In particular, we follow two different strategies to determine the structure of the controlled unitary operation, measure it and analyze it by using machine learning techniques. Here, the resources are quantified by the number of times the unitary operation has to be applied and the output measured in order to be able to determine its structure.

\subsubsection*{Machine Learning}
Here, we employ state-of-the-art classical machine learning algorithms to compare with our quantum protocol. We show the results achieved for three different networks, the two-node line, and two different instances of the three-node line, all of them previously studied with our algorithm in Fig.~$2$. The numerical experiment is designed for determining the optimal control state by evaluating the action of $U$ on the tensor product of a random control state and the fixed $\ket{\text{in}}$. The data consists of a recompilation of random control states, which cover the whole control subspace, with their correspondent fidelity for a fixed $\{ \ket{\text{in}}, \ket{\text{out}} \}$ pair.

 For each network, three data sets were used (small, medium, large) with a different number of instances. It must be emphasized that all results are referred to test sets, i.e., obtained with data not used to train the models. Therefore, they must be taken as a good estimation of the prediction capability of the models for new unseen data. Cross-validation was implemented by means of a $k$-fold approach \cite{Alpaydin}, where $k$=10 for all data sets, except for the small data set of the two-line network whose value was $k$=5 due to the very limited number of instances.

All results were achieved by using Support Vector Regressors (SVRs) \cite{scholkpf}, whose characteristics make them especially adequate when dealing with sparse data sets (few instances and high dimension). SVRs work by creating a transformed data space in which the problem is more easily solvable (ideally the problem is transformed into a linear one). That transformation between spaces is carried out by the so-called kernels (Gaussian and polynomial kernels have been used in this experimentation). The data used for training the models has been randomly selected from a set of multiple pairs of control state and fidelity. Although other ML approaches, such as Reinforcement Learning (RL), might seem appropriate to solve this problem, note that the goal of the problem is actually a prediction of the efficiency of the solution rather than the optimal sequence of steps that link the input state with the output state, thus not matching the RL paradigm

Tables~\ref{2nodes}, \ref{3nodesa} and \ref{3nodesb} report the results achieved by the SVR in the three analyzed networks. These correspond to the two and three node lines analyzed in Fig. 2. In the case of $n=3$, the topology of networks A and B is the same, the one depicted in Fig. \ref{fl1}., but they are defined in a different control basis. For each case, the state with the best fidelity is shown, together with the Mean Error (ME) and the Root Mean Square Error (RMSE). ME is a measure of bias that represents the difference between the real and the predicted efficiencies, i.e., gives information about whether the model tends to make overestimations (negative values) or underestimations (positive values). On the other hand, RMSE is a well-known robust measure of accuracy.

\begin{table}[h!] 
\centering
\begin{tabular}{{|l|l|l|l|l|}}
\hline
Number of Instances  & Small (10) & Medium (75) & Large (500) \\
\hline
  ME & 0.0029 & -1.3 $\times 10^{-4}$ & -8.6 $\times 10^{-5}$  \\
\hline
  RMSE & 0.0493 & 0.0012 & 0.0026 \\
\hline
Best Fidelity & 0.874 & 0.962 & 0.987\\
\hline
\end{tabular}
\caption{{\bf Two-node line.} The optimal control state for this network is $\ket{1}\otimes\ket{1}$, while the best result obtained with this analysis is $(0.0535\ket{0}+0.9986\ket{1})\otimes(0.0786\ket{0}+ 0.9969\ket{1})$.}
\label{2nodes}
\end{table}

\begin{table}[h!] 
\centering
\begin{tabular}{{|l|l|l|l|l|}}
\hline
Number of Instances  & Small (50) & Medium (200) & Large (1000) \\
\hline
  ME & 7.2 $ \times 10^{-4}$ & 2.4 $\times 10^{-5}$ & 3.2 $\times 10^{-4}$  \\
\hline
  RMSE & 0.0054 & 0.0017 & 0.0039 \\
\hline
Best Fidelity & 0.6840 & 0.8836 & 0.8872 \\
\hline
\end{tabular}
\caption{{\bf Three-node line A.} The optimal control state for this network is $\ket{1}\otimes\ket{1}\otimes\ket{1}$, while the best solution that the machine learning protocol provides is $(0.1785\ket{0}+0.9839\ket{1})\otimes(0.2063\ket{0}+ 0.9785\ket{1})\otimes(0.1754\ket{0}+ 0.9845\ket{1})$.}
\label{3nodesa}
\end{table}

\begin{table}[h!] 
\centering
\begin{tabular}{{|l|l|l|l|l|}}
\hline
Number of Instances  & Small (50) & Medium (200) & Large (1000) \\
\hline
  ME & -9.3 $\times 10^{-4}$ & -7.8 $\times 10^{-5}$ & -9.6 $\times 10^{-5}$  \\
\hline
  RMSE & 0.0082 & 0.0018 & 0.0014 \\
\hline
Best Fidelity & 0.9227 & 0.9188 & 0.9709 \\
\hline
\end{tabular}
\caption{{\bf Three-node line B.} The optimal control state for this network is $\frac{1}{\sqrt{2}}[\ket{0}+\ket{1}]\otimes\ket{1}\otimes[\cos{(\pi/3)}\ket{0}+\sin{(\pi/3)}\ket{1}]$, while the result of the analysis is $(0.7512\ket{0}+0.66\ket{1})\otimes(0.1599\ket{0}+ 0.9871\ket{1})\otimes(0.4936\ket{0}+ 0.8697\ket{1})$.}
\label{3nodesb}
\end{table}

\subsubsection*{Measurement of the Unitary Operation}
An alternative method for solving the learning task would be to measure the input-output relation of the controlled unitary operation when strategically, and not randomly, exploring the control subspace. Let us denote by $\ket{c_i}$ the natural basis of the control subspace in $U$, and by $\ket{b_i}$ our guess for this basis in a Hilbert space of dimension $n$. The measurement protocol consists in applying the unitary operation to $\ket{b_i}\otimes\ket{\text{in}}$, projecting this result on $\ket{\text{out}}\bra{\text{out}}$ and tracing out the target subspace achieving $\rho_i$ for each $b_i$. In the worst case, this operation has to be repeated for all $b_i$ to guarantee that the populations of the solutions, and not the internal phases, are found. Afterwards, one has to find the appropriate basis $\ket{c_i}$ as a linear combination of the proposed one $\ket{b_i}$. Another approach is to determine each component of the unitary operation and change to a basis in which the unitary is expressed as a direct sum of the $s_{i}$ operations. This particular strategy highlights the relation between our algorithm and the field of quantum process tomography.  
\\
\subsubsection*{Comparison}
In summary, the purely random approach analyzed with ML techniques requires in principle more resources than the quantum feedback algorithm with delayed equation. Nevertheless, the fact that ML techniques are independent of the basis guarantees their success in any possible situation. The comparison is made between the episodes, the number of times that the time delayed equation has to be repeated, and the instances, the amount of data employed in the ML algorithm. Even if both methods are based on different training mechanisms, the information fed to both of them is the same, a figure of merit for each control state. In the SVR the system is provided with pairs of control state and its correspondent fidelity, which requires the implicit knowledge of $\{\ket{\text{in}}, \ket{\text{out}}\}$ and the ideal $U$ operation. The connection with the quantum algorithm is that the delay term in Eq.~$2$ provides a distance that works in an analogue way as the fidelity in the SVR. Notice that in the quantum algorithm each episode only requires a pair of \{$\ket{\text{in}}$, $\ket{\text{out}}$\} states, therefore the number of episodes equals the number of instances. A more realistic analysis would take into account the duration of each process, but for the moment we cannot make a precise estimation about the time for implementing a time delayed equation.

With respect to the complete measurement approach, recent studies bound its scalability in the order of $n^2$ or even $n$, being the latter the dimension of the Hilbert space \cite{qpt1,qpt2,qpt3}. On the other hand, the measurement protocol does not provide the solution in a physical register, but it is the analysis of the unitary operation that provides the knowledge of it. Moreover, each implementation of the controlled unitary operation is associated with a measurement, while in the quantum machine learning algorithm intermediate measurements are not required, because they are included as an intrinsic part of the dynamics, in contrast to the tomography approach. Additionally,  when measuring, one needs to perform a search for the convenient basis along the Hilbert space to retrieve the correct structure of $U$.

Regarding the scalability of our algorithm, we have observed that the number of episodes for reaching the solution depends on the distance between both, the initial control state and the solution. A direct consequence is that the protocol will not properly work when the initial control state is orthogonal to the solution. This is important to consider because the way to notice the failure is to validate the result by measuring the outcome of the unitary operation. In the simulations carried out here, we have employed $\ket{+}^{\otimes n}$ as the initial control state, but this choice is not unique. In some sense, our protocol can also be understood as a search algorithm. Therefore, a comparison with Grover's result~\cite{yiah} may be in order. Regarding the similarities, the conditional phase rotation in Grover's search algorithm requires the use of an oracle, whose role is played in our formalism by the combination of a controlled unitary operation and the time-delayed terms. On the other hand, the main difference between both protocols is that on Grover's algorithm the basis in which the states to optimize are described is known, while in ours, the search is performed without previous knowledge of the basis, in a similar spirit to the analog algorithm by Farhi and Gutmann \cite{fga}. A positive property of our protocol, in contrast with the previously mentioned quantum search algorithms, is that the solution is reached asymptotically, i.e., the fidelity always increases with the number of episodes.

\section*{Discussion}
In conclusion, we have proposed a quantum machine learning algorithm in which the implementation of time-delayed dynamics allows one to avoid the intermediate measurements, and therefore provides a complementary strategy to conventional quantum machine learning algorithms \cite{ad1,ad2,ad3,ad4}. Moreover, we have shown how the framework of multitask controlled unitary operations is flexible enough to address different problems such as efficient excitation transport in networks. This kind of protocol may be straightforwardly adapted to different quantum architectures, which is beyond the scope of this article. We believe our study represents the first proposal for exploiting feedback-induced effects of delayed-equation dynamics without intermediate measurements in quantum machine learning algorithms.

\section*{Acknowledgements}
The authors acknowledge support from Basque Government grants BFI-2012-322 and IT986-16, Spanish MINECO/FEDER FIS2015-69983-P, Ram\'on y Cajal Grant RYC-2012-11391, and UPV/EHU UFI 11/55.\\

\section*{Author Contribution}
U. Alvarez-Rodriguez, L. Lamata and E. Solano designed the time delayed equation, while P. Escandell-Montero and J. D. Mart\'{i}n-Guerrero analyzed the problem with SVR techniques.\\

\section*{Additional information}
The authors declare no competing financial interests.\\

\end{document}